\documentclass[10pt]{bmc_article}    

\usepackage{cite} 
\usepackage{url}  
\usepackage{ifthen}  
\usepackage{multicol}   
\usepackage[utf8]{inputenc} 
\usepackage{graphics,color}
\usepackage{graphicx}
\usepackage{epsfig}

\newboolean{publ}

\newenvironment{bmcformat}{\baselineskip20pt\sloppy\setboolean{publ}{false}}{\baselineskip20pt\sloppy}

\begin{document}
\begin{bmcformat}


\title{The role of hidden influentials in the diffusion of online information cascades}
 

\author{Raquel A Ba\~nos$^1$%
         \email{Raquel A Ba\~nos\correspondingauthor - raquel.alvarez.ba@gmail.com}
       and 
         Javier Borge-Holthoefer$^1$%
         \email{Javier Borge-Holthoefer - borge.holthoefer@gmail.com}%
       and 
         Yamir Moreno\correspondingauthor$^1$$^2$%
         \email{Yamir Moreno - yamir.moreno@gmail.com}%
      }
      

\address{%
    \iid(1)Institute for Biocomputation and Physics of Complex Systems (BIFI), University of Zaragoza, 50018 Zaragoza, Spain\\
    \iid(2)Department of Theoretical Physics, Faculty of Sciences, University of Zaragoza, 50009 Zaragoza, Spain
}%

\maketitle


\begin{abstract}
In a diversified context with multiple social networking sites, heterogeneous activity patterns and different user-user relations, the concept of ``information cascade'' is all but univocal. Despite the fact that such information cascades can be defined in different ways, it is important to check whether some of the observed patterns are common to diverse contagion processes that take place on modern social media. Here, we explore one type of information cascades, namely, those that are time-constrained, related to two kinds of socially-rooted topics on Twitter. Specifically, we show that in both cases cascades sizes distribute following a fat tailed distribution and that whether or not a cascade reaches system-wide proportions is mainly given by the presence of so-called hidden influentials. These latter nodes are not the hubs, which on the contrary, often act as firewalls for information spreading. Our results are important for a better understanding of the dynamics of complex contagion and, from a practical side, for the identification of efficient spreaders in viral phenomena.
\end{abstract}

\ifthenelse{\boolean{publ}}{\begin{multicols}{2}}{}

\section{Introduction}
Population-wide information cascades are rare events, initially triggered by a single seed or a small number of initiators, in which rumors, fads or political positions are adopted by a large fraction of an informed community. In recent years, some theoretical approaches have explored the topological conditions under which system-wide avalanches are possible \cite{watts2002simple,gleeson2007seed,gleeson2008cascades,hackett2011cascades}; whereas others have proposed threshold \cite{centola2007cascade}, rumor- \cite{borge2012absence} or epidemic-like \cite{leskovec07} dynamics to model such phenomena. Beyond these efforts, digitally-mediated communication in the era of the Web 2.0 has enabled researchers to peek into actual information cascades arising in a variety of platforms --blogs and Online Social Networks (OSNs) mainly, but not exclusively \cite{liben2008tracing,miritello2011dynamical}.

Notably, these latter empirical works deal with a wide variety of situations. First, the online platforms under analyses are not the same. Indeed, we find research focused on distinct social networks such as Facebook \cite{sun2009gesundheit}, Twitter \cite{kwak2010twitter,bakshy2011everyone}, Flickr \cite{cha2009measurement}, Digg \cite{lerman2010information} or the blogospehere \cite{gruhl2004information,adar2005tracking,leskovec07} --which build in several types of user-user interactions to satisfy the need for different levels of engagement between users. As a consequence, although scholars make use of a mostly common terminology (``seed'', ``diffusion tree'', ``adopter'', etc.) and most analyses are based on similar descriptors (size distributions, identification of influential nodes, etc), their operationalization of a cascade --i.e., how a cascade is defined-- largely varies. This fact is perfectly coherent, because how information flows differs from one context to another. Furthermore, even {\em within} the same OSN different definitions may be found (compare for instance \cite{kwak2010twitter} and \cite{bakshy2011everyone}). Such myriad of possibilities is not necessarily controversial: it merely reflects a rich, complex phenomenology. And yet it places weighty constraints when it comes to generalize some results. The study of information cascades easily evokes that of influence diffusion patterns, which in turn has obvious practical relevance in terms of enhancing the reach of a message (i.e. marketing) or for prevention and preparedness. In these applications a unique definition would be highly desirable, as proposed in classical communication theory \cite{rogers1962diffusion}. On the other hand, the profusion of descriptions and the plurality of collective attention patterns \cite{lehmann2012dynamical} hinder some further work aimed to confirm, extend and seek commonalities among previous findings.

In this work we capitalize on a type of cascade definition which pivots on time constraints rather than ``content chains''. Despite the aforementioned heterogeneity, all but one \cite{sun2009gesundheit} empirical works on cascades revolve exclusively around information forwarding: the basic criterion to include a node $i$ in a diffusion tree is to guarantee that (a) the node $i$ sends out a piece of information at time $t_{1}$; (b) such piece of information was received from a friend $j$ who had previously sent it out, at time $t_{2}$; and finally (c) $i$ and $j$ became friends at $t_{3}$, before $i$ received the piece of information (the notion of ``friend'' changes from OSN to OSN, and must be understood broadly here). Note that no strict time restriction exists besides the fact that $t_{1} > t_{2} > t_{3}$, the emphasis is placed on whether the {\em same} content is flowing. This work instead turns to topic-specific data in which it is safely assumed that content is similar, and the inclusion in a cascade depends not on the repetition of a message but rather on the engagement in a ``conversation'' about a matter.

Beyond our conceptualization of a cascade, this work seeks first to test the robustness of previous findings in different social contexts \cite{gonzalez2011dynamics,borge2012locating}, and then moves on towards a better understanding of how deep and fast do cascades grow. The former implies reproducing some general outcomes regarding cascade size distributions, and how such cascades scale as a function of the initial node's position in the network. The latter aims at digging into cascades, to obtain information about their temporal and topological hidden patterns. This effort includes questions such as the duration and depth of cascades, or the relation between community structure and cascade's outreach. Our methodology allows to prove the existence of a subtle class of reputed nodes, which we identify as ``hidden influentials'' after \cite{gonzalez2013broadcasters}, who have a major role when it comes to spawn system-wide phenomena.

\section{Data}
Our data comprises a set of messages publicly exchanged through \textit{www.twitter.com} from the $1^{st}$ of March, 2011, to the $31^{th}$ of March, 2012. The whole sample of messages was filtered by the Spanish start-up company \textit{Cierzo development}, restricting them to those that contained at least one of 20 preselected hashtags (see Table 1). These hashtags correspond to distinct topics, thus we obtained different subsets to which we assign a generic tag. 

We present the results for two of these subsets. One sample consists of 1,188,946 tweets and is related to the Spanish grassroots movement popularly known as ``15M'', after the events on the 15th of May, 2011. This movement has however endured over time, and in this work we will refer to it as {\em grassroots}. Messages were generated by 115,459 unique users. On the other hand, 606,645 filtered tweets referred to the topic ``Spanish elections'', which were celebrated on the third week of November, 2011. This sample was generated by 84,386 unique users. 

Using the Twitter API we queried for the list of followers for each of the users, discarding those who did not show outgoing activity during the period under consideration. In this way, for each data set, we obtain an unweighted directed network in which each node represents an active user (regarding a particular topic). A link from user $i$ to user $j$ is established if $j$ follows $i$. Therefore, out-degree ($k_{out}$) represents the number of followers a node has, whereas in-degree ($k_{in}$) stands for its number of friends, i.e., the number of users it follows. The link direction reflects the fact that a tweet posted by $i$ is (instantaneously) received by $j$, indicating the direction in which information flows. Although the set of links may vary in the scale of months we take the network structure as completely static, considering the topology at the moment of the scrap.

\section{Methods}
\subsection{Time-constrained information cascades}
Twitter is most often {\em exclusively} defined as a microblogging service, emphasizing its broadcasting nature. Such definition overlooks however other facets, such as the use of Twitter to interact with others, in terms of {\em conversations} \cite{honey2009beyond} or {\em collaboration}, for instance connecting groups of people in critical situations \cite{mungiu2009moldova,cha2010measuring}. {\em Addressivity} accentuates these alternative features \cite{honey2009beyond,kwak2010twitter}. Moreover, observed patterns of link (follower relation) reciprocity \cite{garlaschelli2004patterns} (see Table 2) hint further the use of Twitter as an instant messaging system, in which different pieces of information around a topic may be circulating (typically over short time spans) in many-to-many interactions, along direct or indirect information pathways \cite{kossinets2008structure}.

It is precisely in this type of interactions where the definition of a time-constrained cascade is a useful tool to uncover how --and how often-- users get involved in sequential message interchange, in which the strict repetition of contents is not necessary (possibly not even frequent). 
A time-constrained cascade, starting at a \emph{seed} at time $t_{0}$, occurs whenever some of those who ``hear'' the piece of information react to it --including replying or forwarding it-- within a prescribed time frame $(t_{0}, t_{0} + \Delta \tau]$, thereby becoming {\em spreaders}. The cascade can live further if, in turn, reactions show up in $t_{0} + 2\Delta \tau$, $t_{0} + 3\Delta \tau$, and so on. Since messages in Twitter are instantly broadcasted to the set of users following the source, we define listener cascades as those including both active (spreader) and passive participants. In considering so we account for the upper bound of awareness over a certain conversation in the whole population (see Figure~\ref{fig1} for illustration). Admittedly, our conceptualization does not control for exogenous factors which may be occurring at the onset and during cascades. This being so, ours is a comprehensive account of information cascades.

\begin{figure}[tpb]
  \centering
  \includegraphics[width=0.7\columnwidth,clip=0]{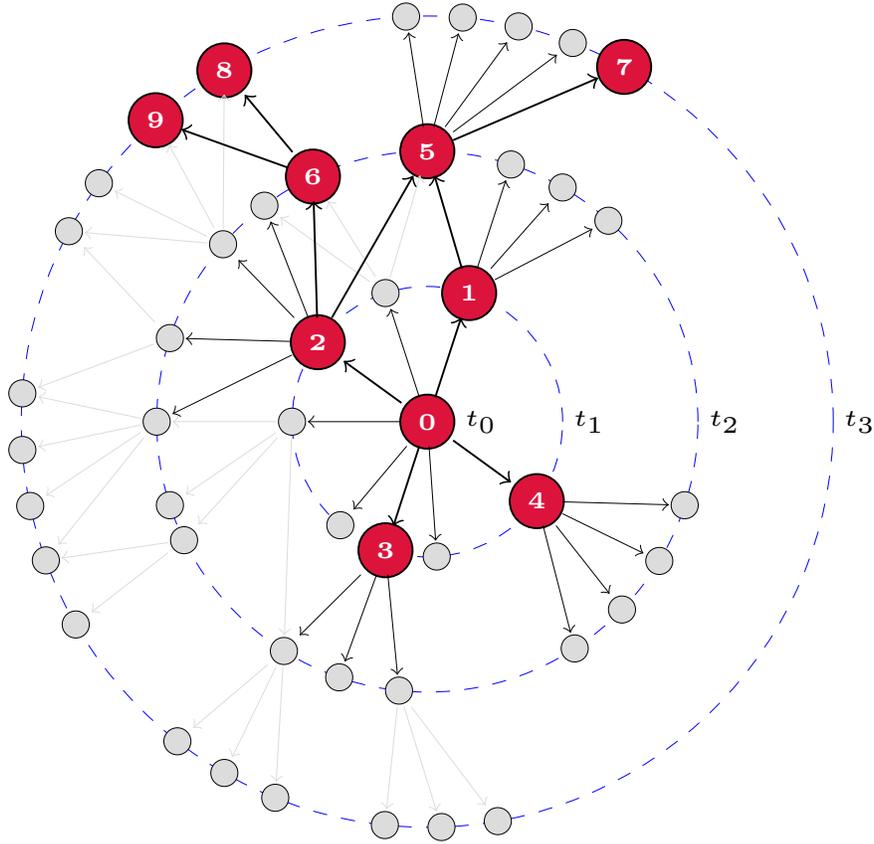}
  \caption{Time-constrained cascades: nodes are disposed in concentric circles indicating the time when they received a specific tweet. Links between them represent the follower/friend relationship: an arrow from $i$ to $j$ indicates that $j$ follows $i$, as any tweet posted by $i$ is automatically received by $j$. Red nodes are those who posted a new message at the corresponding time, whereas gray nodes only \textit{listened} to their friends. In this particular example, user $0$ acts as the initial seed, emitting a message at time $t_0$ which is instantaneously sent to its nearest neighbors, laying on the first dashed circle, who are counted as part of the cascade. Some of them (nodes $1$, $2$, $3$ and $4$) decide to participate at the following time step, $t_1=t_0+\Delta \tau$, posting a new message and becoming intermediate spreaders of the cascade. If any of their followers show activity at $t_2=t_0+2\Delta \tau$ the process continues and the cascade grows in size as new users listen to the message. The process finally ends when no additional users showed activity (as it happens in the cases of users $3$ and $4$), or when an intermediate spreader does not have any followers (users $7$, $8$ and $9$). 
}
\label{fig1}
\end{figure}

We apply the latter definition \cite{gonzalez2011dynamics,borge2012locating} to explore the occurrence of listener cascades in the ``grassroots'' and ``elections'' data. In practice, we take a seed message posted by $s$ at time $t_{0}$ and include all of $s$ followers in the diffusion tree hanging from $s$. We then check whether any of these listeners showed some activity at time $t_{0}+\Delta \tau$, increasing the depth of the tree. This is done recursively, the tree's growth ends when no other follower shows activity. Passive listeners constitute the set of leaves in the tree. In our scheme, a node can only belong to one cascade (but could participate in it multiple times); the mentioned restriction may introduce measurement biases. Namely, two nodes sharing a follower may show simultaneous activity, but their follower can only be counted in one or the other cascade (with possible consequences regarding cascade size distributions or depth in the diffusion tree). To minimize this degeneration, we perform calculations for many possible cascade configurations, randomizing the way we process data. 

In the next sections we report some results for the aforementioned data subsets (``grassroots'', ``elections'') considering all their time span (over one year).
Our results have been obtained for $\Delta \tau = 24$ hours. Previous works \cite{gonzalez2011dynamics,borge2012locating} acknowledge the robustness of cascade statistics for $2 \le \Delta \tau \le 24$; also, a 24-hour window may be regarded as an inclusive bound of the popularity of a piece of information over time on different OSNs, including Twitter \cite{leskovec07,cha2009measurement,sun2009gesundheit,lerman2010information}.

\subsection{Community analyses}

The identification of modules in complex networks has attracted much attention of the scientific community in the last years, and social networks posit a prominent example. A modular view of a network offers a coarse-grained perspective in which nodes are classified in subsets on the basis of their topological position and, in particular, the density of connections between and within groups. In OSNs, this classification usually overlaps with node attribute data, like gender, geographical proximity or ideology \cite{conover2011political,conover2012partisan}.

To detect statistically significant clusters we rely on the concept of modularity $Q$ \cite{newman04b}:
\begin{equation}
Q = \frac{1}{2N_e}\sum_{i}\sum_{j} \left(a_{ij} - \frac{k_{i}k_{j}}{2N_e} \right)\delta(C_{i},C_{j}) 
\end{equation}
where $N_e$ is the number of links in the network; $a_{ij}$ is 1 if there is a link from node $i$ to $j$ and 0 otherwise; $k_{i}$ is the connectivity (degree) of node $i$; and finally the Kronecker delta function $\delta(Ci,Cj)$ takes the value 1 if nodes $i$ and $j$ are classified in the same community and 0 otherwise. Summarizing, $Q$ quantifies how far a certain partition is from a random counterpart (null model).

From the definition of $Q$, algorithms and heuristics to optimize modularity have appeared ever faster and with an increased degree of accuracy \cite{fortunato2010community}. All these efforts have led to a considerable success regarding the quality of detected community structure in networks, and thus a more complete topological knowledge at this level has been attained. In this work we present results for communities detected from the Walktrap method \cite{pons05} in which a fair balance between accuracy and efficiency is sought. The algorithm exploits random walk dynamics. The basic idea behind them is that a random walker tends to get trapped in densely connected parts of the graph, which correspond to communities. Pons and Latapy's proposal is particularly efficient because, as $Q$ is increasingly optimized, vertices are merged into a coarse-grained structure, reducing the computational cost of the dynamics. The resulting clusters at each stage of the algorithm are aggregated, and the process is repeated iteratively.

One of the most useful applications of community analyses is a better understanding of the position of a node \cite{arenas2010optimal}. In terms of information diffusion --and much like in \cite{grabowicz2012social}-- we explore whether community structure (and in particular, the relation of a seed node with the module it belongs to) has an impact on a cascade's success. To do so we adopt the node descriptors proposed by Guimer\`{a} {\em et al.} in \cite{guimera05}: the $z-score$ of the internal degree of each node in its module, and the participation coefficient of a node $i$ ($P_{i}$) defined as how the node is positioned in its own module and with respect to other modules. 

The {\em within-module degree} and the {\em participation coefficient} are easily computed once the modules of a network are known. If $\kappa_i$ is the number of links of node $i$ to other nodes in its module $C_i$, $\overline{\kappa}_{C_i}$ is the average of $\kappa$ over all the nodes in $C_{i}$, and $\sigma_{\kappa_{C_i}}$ is the standard deviation of $\kappa$ in $C_i$, then
\begin{equation}
    z_i=\frac{\kappa_i-\overline{\kappa}_{C_i}}{\sigma_{\kappa_{C_i}}}
\end{equation}
is the so-called {\em z-score}.

The participation coefficient $P_i$ of node $i$ is defined as:
\begin{equation}
    P_i=1-\displaystyle\sum_{C=1}^{N_M}\displaystyle \left({\frac{\kappa_{iC}}{k_i}}\right)^2
\end{equation}
where $\kappa_{iC}$ is the number of links of node $i$ to nodes in module $s$, and $k_i$ is the total degree of node $i$. Note that the participation coefficient $P_i$ has a maximum at $P_i=1-\frac{1}{N_{M}}$, when the $i$'s links are uniformly distributed among all the modules ($N_{M}$), while it is 0 if all its links are within its own module. Those nodes that deviate largely from average internal connectivity are local hubs, whereas large values of $P_i$ stands for connector nodes bridging together different modules.

\section{Results}

\subsection{Cascade size distributions.}
As a starting point, we test the robustness of the results partially presented in \cite{gonzalez2011dynamics}, and further explored in \cite{borge2012locating}. Results shown in Figure~\ref{fig2} confirm these findings. The upper panels show that the size of time-constrained cascades are distributed in a highly heterogeneous manner, with only a small fraction of all cascades reaching system-wide proportions. This is also in good agreement with most preceding works, that have also found that large cascades occur only rarely. On the other hand, when cascades are grouped together such that the reported size corresponds to an average over topological classes, we find that both the degree $k$ (middle panels) and coreness ($k$-core, lower panels) of nodes correlate positively with cascades' sizes. Some theoretical approaches predict similar behavior \cite{kitsak2010identification,borge2012emergence}.

\begin{figure}[]
  \centering
  \includegraphics[width=0.9\columnwidth]{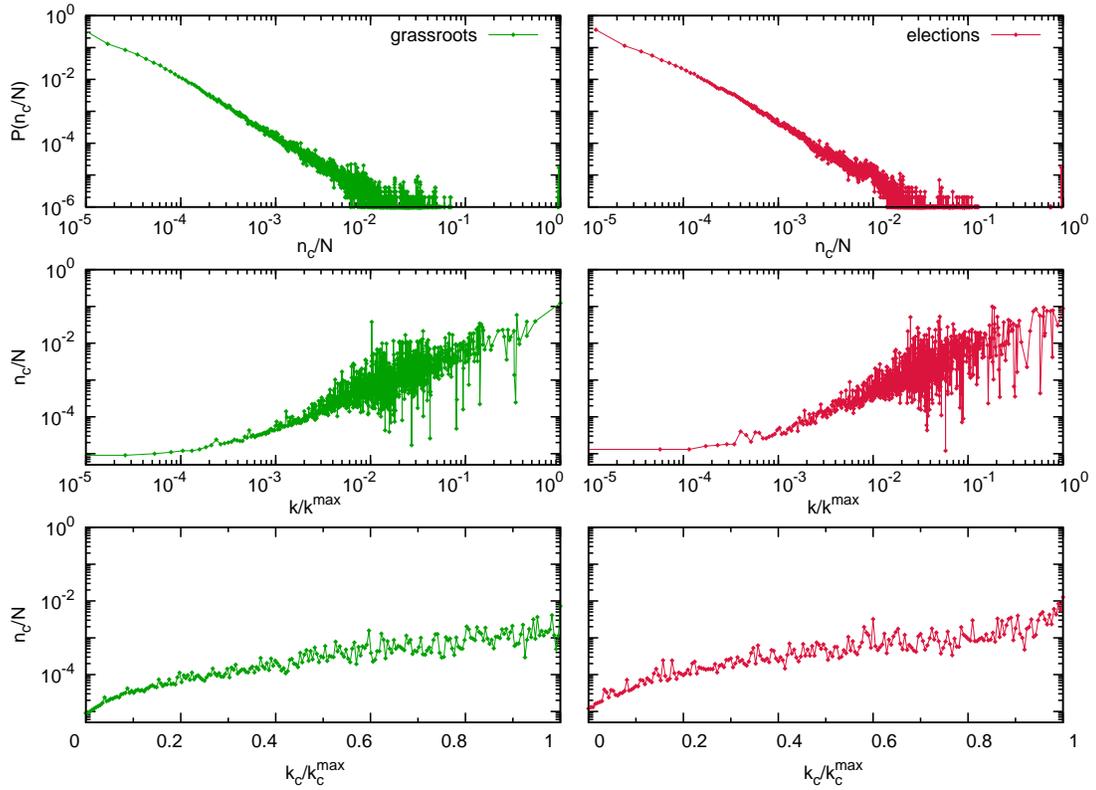}
  \caption{Upper panels: cascade size distributions for the topics under consideration (left: ``grassroots''; right: ``elections''). Middle panels: average spreading capability (rescaled by the system size) grouped by the connectivity of the initial seed, $k$. Lower panels: spreading capability grouped by the $k$-core of the initial spreader. For the sake of comparison, $k$ and $k$-core have been rescaled by their corresponding maxima.}
  \label{fig2}
\end{figure}

\subsection{Cascades' temporal and topological penetration}
The next step to gain insights into the general overview above is to characterize how deep --both temporally and structurally-- a cascade unfolds. We defined the {\em topological penetration}, $\Delta r$ of a cascade as the shortest path between the seed of the cascade and the farthest node involved in the cascade. The results shown in Figure~\ref{fig3} (upper and middle panels) give quantitative support to a well-known fact: most cascades actually die with one single spreader (instantaneous cascades), which corresponds to a shallow tree --though it may be quite wide \cite{leskovec07,kwak2010twitter,bakshy2011everyone}. In this most frequent case, the cascade of listeners simply accounts for the out-degree $k_{out}$ of the seed node. Additionally, the bulk of cascades penetrates up to $\Delta r=3$ or $\Delta r=4$, both for ``grassroots'' and ``elections'', which is in the range of the average path length, but fairly below the upper bound, which is set by the network's diameter (10 and 9 respectively; see Table 2). Interestingly, as shown in the figure, when a cascade moves beyond the average path length between the initial node and any node on the network, namely, to nodes distant $\Delta r > 3$, a large fraction of the population will likely be engaged in a cascade that will reach system-wide sizes with high probability. 

\begin{figure}[]
  \centering
  \includegraphics[width=0.9\columnwidth]{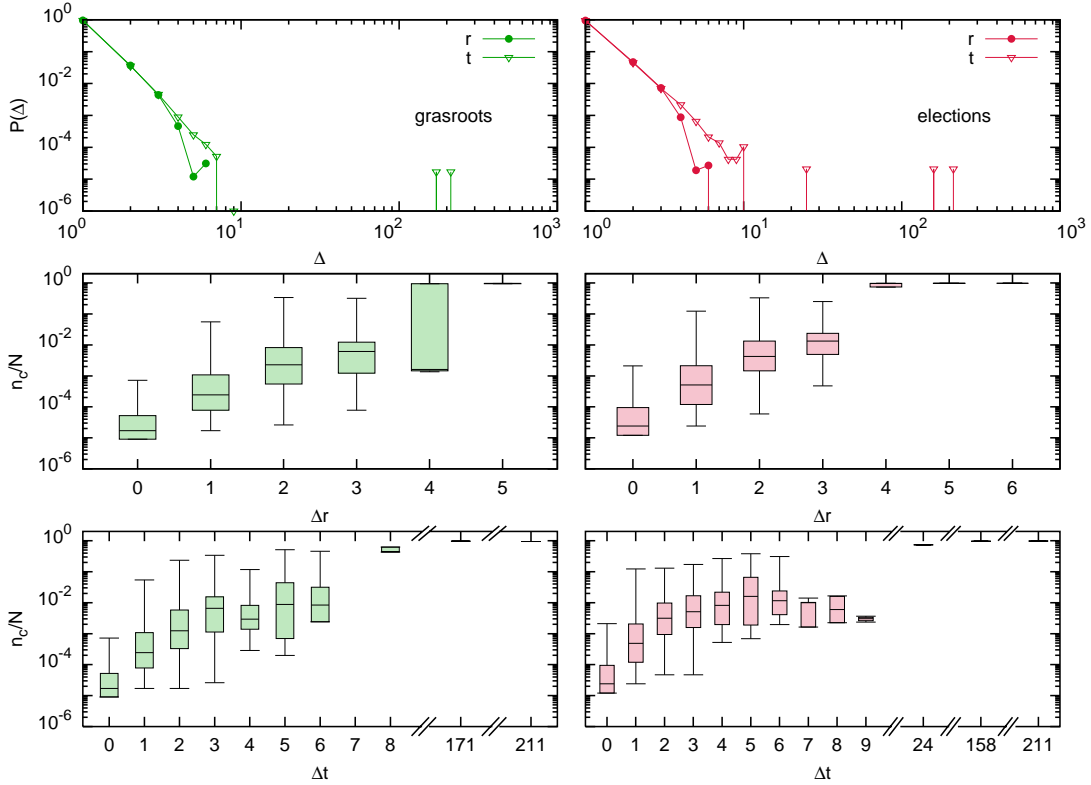}
  \caption{Upper panels: distributions of topological and temporal penetrations. $\Delta r$ is the largest shortest path length between the initial seed and any node involved in the same cascade, where as $\Delta t$ refers to the cascades' lifetimes. Middle panels: box-plots for topological penetration. Lower panels: box-plots for temporal penetration. Cascades' spreading success grows with time, and some exceptional conversations can last for months (note the broken axis).}
  \label{fig3}
\end{figure}

Temporal patterns, as given by the lifetime $\Delta t$ of a cascade, follow similar trends: most cascades die out after 24 hours, which closely resembles previously reported results \cite{leskovec07}. However, in Figure~\ref{fig3} (upper panels) we observe a richer distribution (compared to topological penetration) such that cascades may last over 100 days, suggesting that the survival of a conversation does not exhibit an obvious pattern. Again, this result confirms --from a different point of view-- empirical results published elsewhere \cite{sun2009gesundheit,lehmann2012dynamical}. Finally, temporal penetration illustrates the fact that a node may participate multiple times in a single cascade --although it is counted just once. This is implicit in the definition of a time-constrained cascade, placing it closer to neuronal dynamics and spike-trains --which comprehend self-sustained activity-- and deviating it from classical modeling approaches --such as rumor spreading dynamics, where multiple exposures to the rumor end up in ceasing its dissemination. In any case, Figure~\ref{fig3} (lower panels) illustrates that survival can not guarantee system-wide cascades, although an increasing pattern is observed as survival time grows.



\subsection{Identification and role of hidden influentials}
Up to now we have related a cascade's size to certain features of the seed node. Although we observe a clear positively correlated pattern (the larger the seed's descriptor, the larger the resulting cascade), one might fairly argue that a wide range of values below the maximum produces a similar outcome. So, for instance, seeds in the range $10^{-2} \le k/k^{max} \le 10^{-1}$ (Figure~\ref{fig2}) can sometimes trigger large cascades; the same can be said for $k_{c}/k_{c}^{max} \ge 0.6$. This finding prompts us to hypothesize that the success of an activity cascade might greatly depend on intermediate spreaders characteristics, and not only on the properties of the seed nodes. That being so, a large seed $k_{out}$  (i.e. its follower set) may be a sufficient but not a necessary condition for the generation of large-scale cascades. In this section we explore how the average connectivity of the train of spreaders involved in a cascade affects its final size.

\begin{figure}[]
  \centering
  \includegraphics[width=0.9\columnwidth]{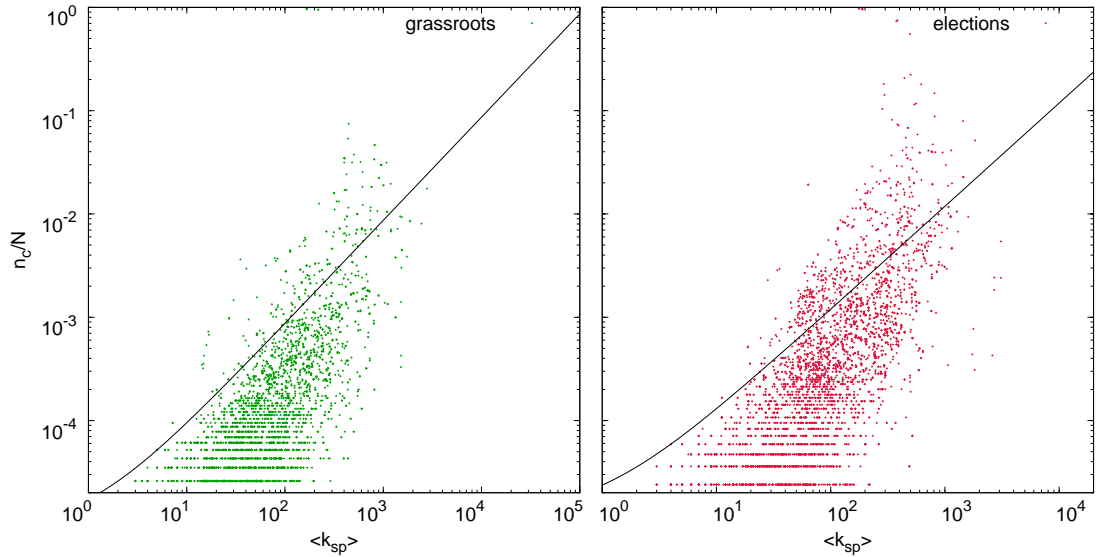}
  \caption{Cascade size {\em vs.} average connectivity of intermediate spreaders $\langle k_{sp} \rangle$. Non-instantaneous cascades are displayed, where the initial seed and its inactive listeners have been removed in order to dismiss the effect of the initial seed on the cascade size. There is a clear correlation between both magnitudes, although some unexpected behavior shows up: the existence of cascades containing ``hidden spreaders'', users who are capable of generating large cascades despite not having hub-like connectivity. In both panels the function $y=(x+1)/N$ is drawn as a reference. 
  }
  \label{fig4}
\end{figure}

To study the role of intermediate spreaders we split our results, distinguishing instantaneous cascades (those with a unique spreader) from those with multiple spreaders. The former merely underlines the fact that the seed's $k_{out}$ suffices to observe large cascades. The latter, more interestingly, unveils a new character in the play: {\em hidden influentials}, relatively smaller (in terms of connectivity) nodes who, on the aggregate, can make chain reactions turn into global cascades. Figure~\ref{fig4} reveals these special users: note that the largest effects are obtained for those spreaders who, on average, have $10^{2}$ to $10^{3}$ neighbors (both for ``grassroots'' and ``elections''). These nodes do not occupy key topological positions that would {\em a priori} identify them as influential, and yet they play a major role promoting system-wide events \cite{watts07influentials,gonzalez2013broadcasters}. Therefore, getting these nodes involved has a multiplicative impact on the size of the cascades. 

To quantify such effect, we introduce the {\em multiplicative number} of a given node $i$, $\Delta l$ (in analogy with the basic reproductive number in disease spreading), which is the quotient of the number of listeners reached one time step after $i$ showed activity, $l\left(t+1\right)$, and the number of $i$'s nearest listeners, i.e., those who instantaneously received its message, $l\left(t\right)$ (which is given by the number of followers of $i$ that are involved in the cascade). Thus, the ratio $\Delta l$ measures the multiplicative capacity of a node: $\Delta l>1$ indicates that a user has been able to increase the number of listeners who received the message beyond its immediate followers. 

Figure~\ref{fig5} shows how $\Delta l$ is distributed as a function of $k_{out}$. Top panels represent the proportion of nodes with $\Delta l>1$ and $\Delta l \le 1$ per degree class. In this case, normalization takes into account all possible $k_{out}$ and all $\Delta l$ (above and below 1) counts, so as to evidence  that in most cases cascades become progressively shrunk as they advance. The fact that the area corresponding to the region $\Delta l>1$ is much smaller than that for $\Delta l \le 1$ tells us that most cascades are small, which is consistent with the reported cascades' size distribution. On the other hand, bottom panels in Figure~\ref{fig5} focus on the same quantity, but in this case we represent the probability $P(k_{out},\Delta l > 1)$ ($P(k_{out},\Delta l \le 1)$) that a node of out-degree $k_{out}$ has (does not have) a multiplicative effect. As before, the results indicate that, in both datasets, the most-efficient spreaders (those with a multiplicative number larger than one) can be found most often in the degree classes ranging from $k_{out}=10^{2}$ to $k_{out}=10^{3}$, i.e., significantly below $k_{max}$ (see Table 2). These nodes are the actual responsible that cascades go global and must be engaged if one would like to increase the likelihood of generating system-wide cascades. 

\begin{figure}[]
  \centering
  \includegraphics[width=0.9\columnwidth]{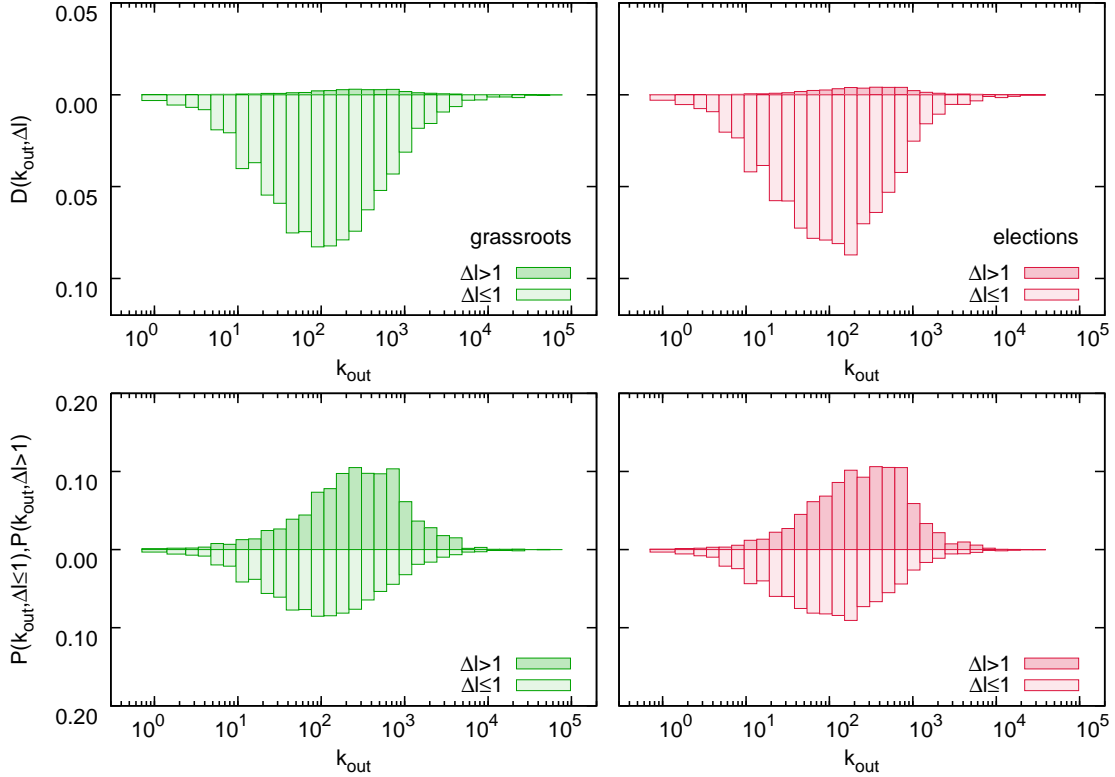}
  \caption{Top panels represent the distribution of the multiplicative number, $\Delta l$ for both datasets as indicated. As one can see, most of the time $\Delta l \le 1$, which reflects the rare occurrence of large cascades. The bottom panels instead represent the probabilities, $P(k_{out}, \Delta l > 1)$ (above 0 baseline) and $P(k_{out}, \Delta l \le 1)$ (below it) that a node of out-degree $k_{out}$ has or does not have a multiplicative effect, respectively. See the text for further details.}
  \label{fig5}
\end{figure}

The previous features of hidden influentials poses some doubts about what is the actual role of hubs in cascades that are not initiated by them. Interestingly, we next provide quantitative evidences that, in contrast to what is commonly assumed, hubs often act as cascade firewalls rather than spawners. To this end we have measured $\langle k_{nn} \rangle$ (average nearest neighbors degree) with respect to seed nodes. Each point in Figure~\ref{fig6} represents the relationship between cascade size and $\langle k_{nn} \rangle$. The initial trend is clear and expected: the larger is the average degree of the seed's neighbors, the deeper the tree grows. However, at some point this pattern changes and indicate that cascades may die out when they encounter a hub, more often than not. If this were not the case, one would observe a monotonically increasing dependence with $\langle k_{nn} \rangle$. This counterintuitive hub-effect is mirrored in classical rumor dynamics \cite{borge2012absence} and can be explained scrutinizing the typically low activity patterns of these (topologically) special nodes \cite{cha2010measuring,borge2011structural}.

\begin{figure}[]
  \centering
  \includegraphics[width=0.9\columnwidth]{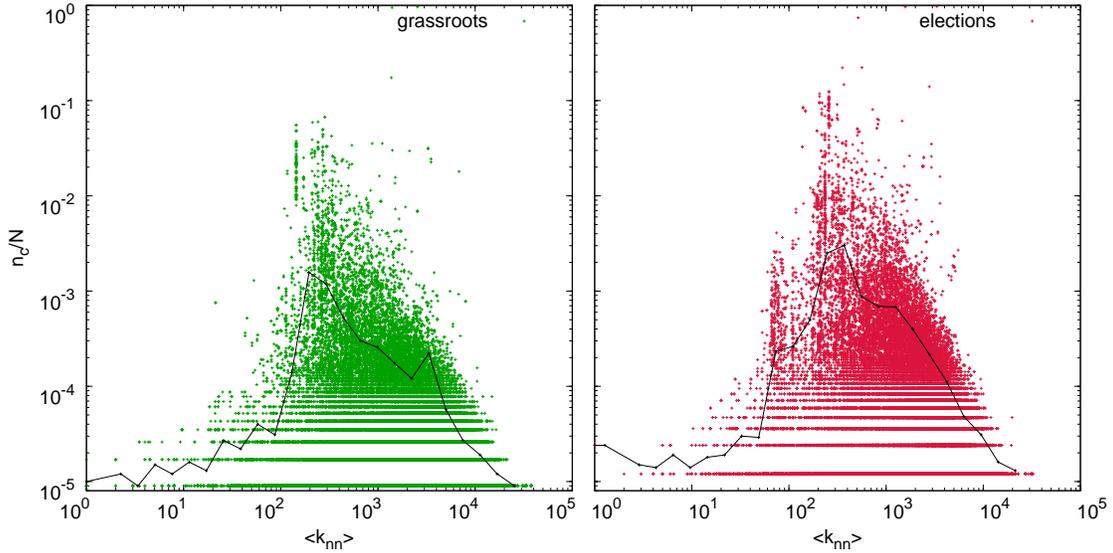}
  \caption{Cascades' outreach as a function of $\langle k_{nn} \rangle$ (average nearest neighbors degree): remarkably, nodes with the highest connectivity do not enhance, but rather diminish, cascades growth. As in the previous Figure, largest cascades are obtained when second spreaders (the seed's neighbors) have, on average, $10^{2} < k < 10^{3}$.}
  \label{fig6}
\end{figure}

\subsection{The role of community structure in information diffusion}
It is generally accepted that cohesive sub-structures play an important role for the functioning of complex systems, because topologically dense clusters impose restrictions to dynamical processes running on top of the structure \cite{arenas2006synchronization,danon2008impact}. For example, in the context of OSNs, detected communities in {\em @mention} Twitter networks were found to encode both geographical and political information \cite{borge2011structural}, suggesting that a large fraction of interactions take place locally, but lots of them also correspond to global modules --for instance, users rely on mass media accounts to amplify their opinion. Focusing on information diffusion, inter- and intra-modular connections in OSNs have already been explored \cite{grabowicz2012social} regarding the nature of user-user ties. We instead investigate other questions, such as: (i) are modules actual bottlenecks for information diffusion?; (ii) is the spreading of information more successful for ``kinless'' nodes (those who have links in many communities besides their own one)? Or (iii) do local hubs --those with larger-than-expected intra-modular connectivity-- have higher chances to trigger system-wide cascades?

\begin{figure}[]
  \centering
  \includegraphics[width=0.9\columnwidth]{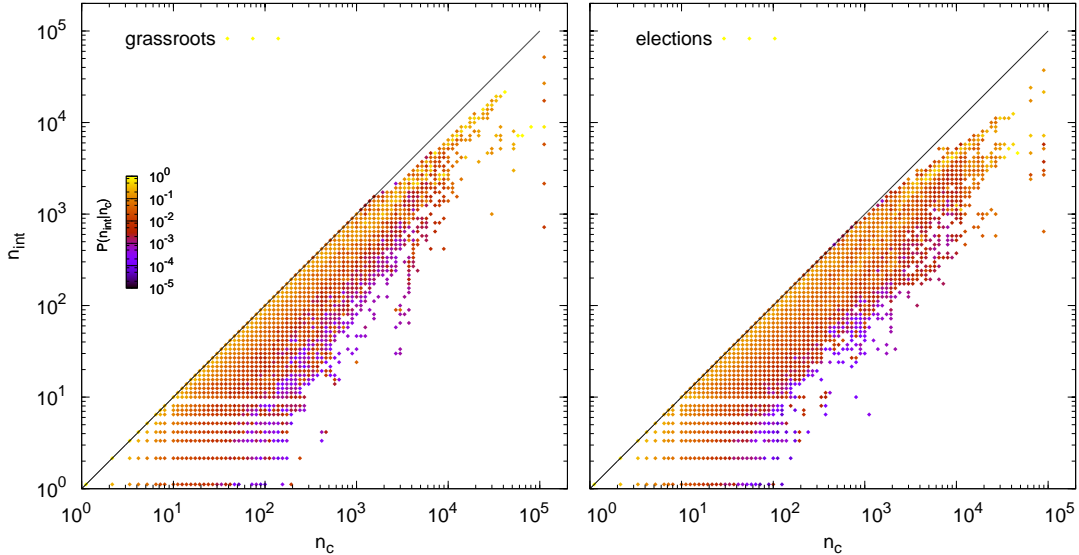}
  \caption{Inter- and intra-modular cascading events. Binned representation of the relationship between the number of nodes ($n_{int}$) in a cascade that unfolds in the same community of the initial seed and the size of the cascade itself ($n_{c}$). Proportions have been normalized column-wise, i.e. by the total number of cascades with same size. Note that cascades affecting up to $10^{3}$ nodes mostly lie close to the diagonal, i.e. a vast majority of the cascade occurs within the community where it began. At a certain point (beyond $n_{c} \approx 10^{3}$) cascades spill over the module where they began.}
  \label{fig7}
\end{figure}

We apply the community analyses described in Section 3.2 and obtain a network partition in $S=5,838$ and $S=4,665$ modules, for the ``grassroots'' and ``elections'' data sets respectively, with optimized $Q$ values and maximum module size $S_{max}$ given in Table 2. Next, for each cascade we compute how many nodes in the resulting diffusion tree belong to the same cluster of the seed ($n_{int}$). This allows to get, as shown in Figure~\ref{fig7}, how often a cascade spills over the module where it began. Interestingly, small to medium-sized cascades ($\sim 10^3$) mainly diffuse within the same community where they were prompted, which hints at the fact that influence occurs within specialized topics \cite{cha2010measuring}. Note however that our approach to community analysis is blind to contents and relies solely on the underlying topology, thus we can only make an educated guess regarding whether modules cluster users around a certain topic (i.e. assuming {\em homophily} \cite{mcpherson2001birds,centola2007homophily}). Remarkably, our results match --qualitatively at least-- the predicted behavior in \cite{gleeson2008cascades} regarding cascades in correlated and modular networks.

Turning to the individual level, the results depicted in the $z-P$ plane of Figure~\ref{fig8} confirm the importance of connectivity --in this case, within-module leadership-- to succeed when a cascade is triggered. Indeed, most nodes for which $z>1$ elicit large cascades in both samples. However, and most interestingly, it suggests that connector or kinless ($P_{i} > 0.6$) nodes \cite{guimera05} can perform better than expected at precipitating system-wide cascades just by paying attention to internal connectivity. As shown in the figure, nodes with a z-score between 0 and 1 acting as connectors are still able to generate system-wide cascades because they compensate their relative lack of connectivity by bridging different modules. This feature is specially noticeable in the case of the ``election'' dataset (right panel). All in all, our results establish that topological modules represent indeed dynamical bottlenecks, which need to be bypassed --through high but also low connectivity users-- to let a cascade go global.

\begin{figure}[]
  \centering
  \includegraphics[width=0.9\columnwidth]{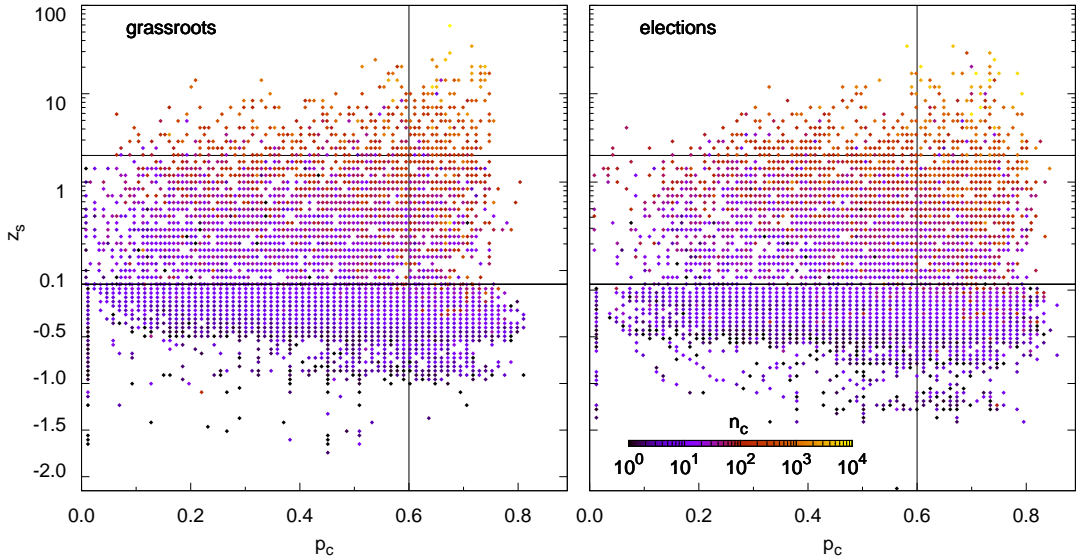}
  \caption{Color-coded $z-P$ planes to assess whether the modular structure of the following/friend network places dynamical constraints in the growth of cascades. A first clear result is that when local leaders ($z > 2$) precipitate information cascades, these tend to be more successful. More interestingly, connector nodes ($P_{i} > 0.6$) also succeed quite often, suggesting that a node's position in the mesoscale can sometimes play a more important role than a rich connectivity.}
  \label{fig8}
\end{figure}

\section{Discussion}

In just one decade social networking sites have revolutionized human communication routines, placing solid foundations to the advent of the Web 2.0. The academia has not ignored such eruption, some researchers foreseeing a myriad of applications ranging from e-commerce to cooperative platforms; while others soon intuited that OSNs could represent a unique opportunity to bring empirical evidence at large into open sociological problems. Information cascades fall somewhere in between, both attracting the interest of viral marketing experts --who worry about optimal outreach and costs-- and collective social action and political scientists --concerned about grassroots movements, opinion contagion, etc.

However, the diversity of OSNs --which constrains the format and the way information flows between users-- and the complexity of human communication patterns --heterogeneous activity, different classes of collective attention-- have resulted in a multiplicity of empirical approaches to cascading phenomena --let alone theoretical works. While all of them highlight different interesting aspects of information dissemination, little has been done to confirm results testing its robustness across different social platforms and social contexts. 

In this regard, the present work capitalizes on previous research to collect, in new large datasets, the statistics of time-constrained information cascades. This scheme exploits the concept of spike train from neuroscience, i.e., a series of discrete action potentials from a neuron. In the brain, two regions are classified as functionally related if they show activity within the same time window. Consequently, message chains are reconstructed assuming that conversation-like activity is contagious if it takes place in relatively short time windows. The main preceding observed trends are here reproduced successfully. Furthermore, we extend the study to uncover other internal facets of these cascades. First, we have discussed how long in time and how deep in the topology cascades go, to realize that, as in neuronal activity, time-constrained cascades can exhibit self-sustained activity. We have then paid attention to those nodes who, beyond the seeds initial onset, actively participate in the cascade. 

Our main results point to two counterintuitive facts, by which hubs can short-circuit information pathways and average users --hidden influentials-- spawn system-wide events. We have found that for a cascade to be successful in terms of the number of users involved in it, key nodes should be engaged. These nodes are not the hubs, which more than often behave as firewalls, but a middle class that either have a high multiplicative capacity or act as bridges between the modules that make up the system. Presumably, modular topologies --abundant in the real world-- entail the presence of information bottlenecks (poor inter-modular connectivity) which place constraints to efficient diffusion dynamics. Indeed, we find that medium-sized and small cascades (the most frequent ones) happen mainly within the community where a cascade sprung. Furthermore, those seed nodes which happen to be poorly classified (they participate in many modules besides their own) are more successful at triggering large cascades.

A better understanding of time-constrained cascading behavior in complex systems leads to new questions. First, it seems clear that the bulk of theoretical work devoted to information spreading is not meant to model this conversation-type dynamics --it is rather focused on rumor and epidemic models. Other approaches need to be sought to fill such gap. Also, time-constrained cascades have always been studied in the context of political discussion and mobilization. As such, this is a fairly limited view of what happens in a service with (as of late 2012) over 200 million active users. Results like the ones obtained here will anyhow provide new hints for a better understanding of social phenomena that are mediated by new communication platforms and for the development of novel manmade algorithms for effective and costless dissemination (viral) dynamics.

\bigskip

\section*{Acknowledgements}
  \ifthenelse{\boolean{publ}}{\small}{}
 We thank Dr. A Rivero for helping us to collect and process the data used in this paper. This work has been partially supported by MINECO through Grant FIS2011-25167; Comunidad de Arag\'on (Spain) through a grant to the group FENOL and by the EC FET-Proactive Project PLEXMATH (grant 317614).
 
\section*{Competing interests}
The authors declare that they have no competing interests.

\section*{Author's contributions}
RAB, JBH and YM conceived the experiments. RAB and JBH performed the analysis.  All authors wrote and approved the final version of the manuscript.
 

\newpage
{\ifthenelse{\boolean{publ}}{\footnotesize}{\small}
 \bibliographystyle{bmc_article}  
  \bibliography{biblio} }     

\ifthenelse{\boolean{publ}}{\end{multicols}}{}



\section*{Tables}

\subsection*{Table 1 - Filtered hashtags and keywords}
Both ``grassroots'' and ``elections'' data sets were collected filtering Twitter traffic according to related keywords, which are listed in this table. For each keyword we display the number of hashtags found (keywords preceded by '\#'), the number of mentions (keywords preceded by '@') and the number of words (keywords with no preceding symbol).

    \par \mbox{}
    \par
    \mbox{
    
\begin{tabular}{llrrr}
\label{tab1}

Keyword & Topic & Hashtags & Mentions & Words \\
\hline
 15m                        & grassroots &      389818      &    3475  &   132049 \\
 acampada             & grassroots &        13732      &    3423  &    76689 \\
 acampadasol        & grassroots &       251344      &   90737  &     3866 \\
 anonymous            & grassroots &        70037      &    4188  &   112859 \\
 democraciarealya & grassroots &        81256      &    1893  &     8798 \\
 indignados             & grassroots &        23371      &     348  &   185615 \\
 nonosvamos          & grassroots &        63490      &     124  &      245 \\
 notenemosmiedo  & grassroots &        35249      &     106  &       55 \\
 occupy                    & grassroots &        18223      &    1467  &    39037 \\
 perroflauta              & grassroots &         1394      &      20  &    26325 \\
 spanishrevolution  & grassroots &       242426      &     926  &     3123 \\
\hline
 20n              &elections &    180323 &         227 &     71440 \\
 25m             &elections &     59812  &          40 &     11887 \\
 elecciones  &elections &     30935  &         269 &    593046 \\
 hondt           &elections &         5      &           0 &      3713 \\
 iu                  &elections &      2726  &        1127 &     33168 \\
 nolesvotes  &elections &    156133 &        2984 &      4621 \\
 pp                &elections &     20412  &        3106 &    201136 \\
 psoe            &elections &     14896  &       22681 &    122222 \\
 vota             &elections &     11464   &         297 &    246764 \\

\end{tabular}
      }

\subsection*{Table 2 - Network properties summary}
$N_v$ number of vertices, and $N_e$ number of edges. WCC stands for the size of the weakly connected component; SCC is the size of the strongly connected component.
Next we report the maximum degree and core values for the undirected network ($k_{in}+k_{out}$), network of friends ($k_{in}$), and network of followers ($k_{out}$).
Average shortest path $L$ and diameter $D$ (the largest shortest path in the network) provide some hints about how deep in the structure can a cascade travel.
Reciprocity $\rho$ is a type of correlation expressing the tendency of vertex pairs to form mutual connections. Notably, results for the datasets in this works are higher than those for social networks in \cite{garlaschelli2004patterns}, and are actually comparable to reciprocity in neural networks. In our context, it reinforces the idea that Twitter may be used {\em both} as a microblogging system and a message interchange service.
Community detection parameters. Louvain algorithm (L) and radatools (RT) with an extremal optimization heuristic (e) and fast-algorithm (f) have been used for comparison. $Q$ stands for the best modularity found and $N_{com}$ for the number of communities detected. The quotient of the largest community's size and the network size, $S^{max}/N$, is also shown.

    \par \mbox{}
    \par
    \mbox{
    
\begin{tabular}{lrr}
\label{tab2}
 & grassroots & elections\\[4pt]
\hline
\hline
Network descriptor & & \\
\hline
$N_v$ & 115,459 & 84,386\\[2pt]
$N_e$ & 10,191,105 & 7,427,825\\[2pt]
WCC & 113,671 & 83,331\\[2pt]
SCC & 102,750 & 76,941\\[2pt]
max($k_{in}+k_{out}$) & 38,028 & 32,073\\[2pt]
max($k_{in}$) & 8,262 & 7,924\\[2pt]
max($k_{out}$) & 37,810 & 31,402\\[2pt]
max($k_{c}$) (undirected) & 228 & 210\\[2pt]
L & 3.175 & 3.092\\[2pt]
D & 10 & 9\\[2pt]
r & -0.116 & -0.124\\[2pt]
$\rho$ & 0.455 & 0.489\\[5pt]
\hline
Mesoscale characterization & & \\[2pt]
\hline
$Q$ & 0.413 & 0.448\\[2pt]
$N_{com}$ & 5,838 & 4,665\\[2pt]
$S_{max}/N$ & 0.196 & 0.110\\
\end{tabular}
      }

\end{bmcformat}
\end{document}